\documentstyle[12pt]{article}
\begin{document}
\baselineskip 24pt
\bibliographystyle{unstr}
\vbox{\vspace{6mm}}
\begin{center}
{\large \bf Causality, Memory Erasing and Delayed Choice Experiments}
\\[9mm]
  Y. Aharonov$^{a,b}$, S. Popescu$^a$ and L. Vaidman$^a$\\[8mm]
{\it   $^a$ School of Physics and Astronomy
\\Raymond and Beverly Sackler Faculty of Exact Sciences
\\ Tel--Aviv University,
\ \ Tel--Aviv \ \ 69978 \ \ ISRAEL\\[5mm]
 $^b$ Physics Department \\ University of South Carolina \\Columbia,
South Carolina 29208 }
\end{center}
\vspace{2mm}

Recently, Ingraham \cite{ING} suggested ``a delayed-choice experiment with
partial, controllable memory erasing''.  We shall show that he cannot be right
since
 his predictions contradict relativistic causality. We shall explain a
subtle quantum effect which was overlooked by Ingraham.

Let us first sketch  Ingraham's argument. He  considered two
atoms located close to each other. The atoms have four relevant levels: the
ground state
$|c\rangle$ and three
excited levels, $|a\rangle, |b\rangle, |b'\rangle$. The first laser pulse
puts the atoms into the superposition ${1\over \sqrt 2}( |a_1 c_2\rangle +
|c_1 a_2\rangle)$.  The atom in the state $|a\rangle$ immediately
emits a photon $\gamma$ and ends up in the state $|b\rangle$.
Then the quantum evolution of  the system  is:
\begin{equation}
{1\over \sqrt 2}( |a_1 c_2\rangle + |c_1 a_2\rangle) \rightarrow
 {1\over \sqrt{2}}(|b_1 c_2\rangle  |\gamma_1\rangle + |c_1 b_2\rangle
|\gamma_2\rangle)
\end{equation}
 We are looking
for an interference between the states of the photon $\gamma$ emitted from
the two atoms on a distant screen.  Since
there is a ``memory'' in the atoms of which atom emitted the photon (the state
$|b\rangle$), no
interference pattern is expected.

Now we add a second pulse, after the first one, but before the photon
$\gamma$ reaches the screen. The second pulse excites the atom in  the state
$|b\rangle$ to the state $|b'\rangle$ which immediately decays to the
ground state $|c\rangle$ and emits  a photon
$\phi$.
In the limiting case, when the wavelength of the photon $\phi$ is much
larger than the distance between
the atoms, the states $\phi_1$ and $\phi_2$ of the photon  emitted from the two
atoms are
practically identical,  $|\phi_1\rangle \simeq
|\phi_2\rangle \equiv |\phi\rangle$.
Therefore, the memory of which atom emitted the
photon $\gamma$ is erased. Indeed, the quantum evolution is
\begin{equation}
 {1\over \sqrt{2}}(|b_1 c_2\rangle  |\gamma_1\rangle + |c_1 b_2\rangle
|\gamma_2\rangle) \rightarrow
{1\over \sqrt{2}}(|b'_1 c_2\rangle  |\gamma_1\rangle + |c_1 b'_2\rangle
|\gamma_2\rangle) \rightarrow
{1\over \sqrt{2}} |c_1 c_2\rangle |\phi\rangle ( |\gamma_1\rangle +
|\gamma_2\rangle)
\end{equation}
 Consequently, in the experiment with
these two laser pulses we expect to  see an interference pattern on the
screen.

If the above analysis  were correct we could send signals faster than
light in the following way. Consider an experiment performed on an ensemble of
such pairs of atoms. Bob, who has to receive the signal, is prepared to measure
the interference pattern on the screen. Alice, who is located near the atoms,
decides
to apply or not to apply the second laser pulse.  Bob will see the interference
pattern, only if Alice applied the second
pulse. Thus, Bob will know  Alice's decision with superluminal velocity.

The argument of Ingraham is based on the fact that before the second laser
pulse the two states $ |b_1 c_2\rangle $ and $|c_1 b_2\rangle $ are practically
orthogonal, while after the second pulse they
evolve into the states $ |c_1 c_2\rangle  |\phi_1\rangle$ and $
|c_1 c_2\rangle  |\phi_2\rangle$ which are practically identical
since $|\phi_1\rangle$, the state of the photon emitted by the first atom
is almost identical  to $|\phi_2\rangle$, emitted by the second atom. This
process  violates  the  unitarity of quantum theory and, therefore, it
cannot take place.

But where is the mistake? The resolution is somewhat subtle. It is true
that a single  atom in the state $|b'\rangle$ immediately radiates photon
$\phi$.
However,  if we put another atom
in the ground state very close to the first one (such that the scalar product
of
the states of the photon emitted from the two locations approximately equals
1), the probability to emit
the photon reduces to 1/2 \cite {DICK}.  Indeed, while in the symmetric
state $|\Psi_+\rangle = {1\over \sqrt 2} (|b'_1 c_2\rangle  + |c_1
b'_2\rangle)$ the atoms emit the photon $\phi$ immediately, in the
antisymmetric state $|\Psi_{-}\rangle = {1\over \sqrt 2} (|b'_1 c_2\rangle  -
|c_1 b'_2\rangle)$ the atoms can not  emit the photon.
 One can understand this phenomenon as a destructive interference between
the states of the photon emitted by the two atoms in the antisymmetric
state.
The correct evolution, instead of Eq. (2)  is:
\begin{eqnarray}
 {1\over \sqrt{2}}( |b'_1 c_2\rangle  |\gamma_1\rangle + |c_1 b'_2\rangle
|\gamma_2\rangle)=
 {1\over 2} \Bigl((|\Psi_+\rangle + |\Psi_-\rangle)|\gamma_1\rangle +
(|\Psi_+\rangle -  |\Psi_-\rangle)|\gamma_2\rangle \Bigr) \rightarrow
\nonumber \\
 {1\over 2} \Bigl((|c_1 c_2\rangle  |\phi \rangle +
|\Psi_-\rangle)|\gamma_1\rangle +
(|c_1 c_2\rangle  |\phi \rangle -  |\Psi_-\rangle)|\gamma_2\rangle  \Bigr)
\end{eqnarray}
Since the states ${1\over \sqrt 2}(|c_1 c_2\rangle |\phi \rangle +
|\Psi_-\rangle)$ and
${1\over \sqrt 2}(|c_1 c_2\rangle  |\phi \rangle - |\Psi_-\rangle)$ are
orthogonal, there is no  memory
erasing.
(Since the states $ |\phi_1 \rangle$ and  $|\phi_2
\rangle$ are not exactly orthogonal,  the state
$|\Psi_-\rangle$  is actually only a metastable state and it  will finally
decay emitting a photon $\phi '$ . However, since it
can be emitted only long  after the emission time of $\phi $, the
 state $|\phi '\rangle$ is practically
orthogonal to $|\phi \rangle$. Thus, after the final decay,  the memory is
stored in the photon
state.)

One may suspect that since the photon states $ |\gamma_1\rangle$ and
$|\gamma_2\rangle$   are obtained by emission from the same very close
atoms, similar phenomena will happen when the photon $ \gamma$ is emitted.
This, however, is not the  case: In order to  be able to yield  an
interference pattern the wavelength of $ \gamma$ has to be taken  smaller
than the distance between the atoms and  therefore the photon states  $
|\gamma_1\rangle$ and
$|\gamma_2\rangle$  are almost orthogonal.

\end{document}